%% file: main.tex
\documentclass[sigconf]{acmart}

\copyrightyear{2018}
\acmYear{2018}
\setcopyright{rightsretained}
\acmConference[MIS2]{WSDM workshop on Misinformation and Misbehavior Mining on the Web}{2018}{Marina Del Rey, CA, USA}
\acmDOI{10.475/123_4}
\acmISBN{123-4567-24-567/08/06}

\usepackage[english]{babel}
\usepackage[utf8x]{inputenc}
\usepackage[T1]{fontenc}

\usepackage{amsmath}
\usepackage{graphicx}
\usepackage[colorinlistoftodos]{todonotes}

\usepackage[tight,footnotesize]{subfigure}
\usepackage{xspace}
\usepackage{threeparttable}
\usepackage{array}

\newcommand{\rh}{highly-active hour\xspace}

\newcommand{\rhs}{highly-active hours\xspace}
\newcommand{\Rhs}{Highly-active hours\xspace}

\newcommand{\rcmm}{reported comment\xspace}
\newcommand{\rcmms}{reported comments\xspace}

\newcommand{\rusers}{reported users\xspace}

\newcommand{\cmms}{users\xspace}

\newcommand{\lub}{latent behavior\xspace}
\newcommand{\Lub}{Latent behavior\xspace}
\newcommand{\lubs}{latent behaviors\xspace}

\newcommand{\nrep}{\phi}    
\newcommand{\ncomm}{r}    
\newcommand{\ncommThr}{\rcmms threshold\xspace} 
\newcommand{\VReported}{R} 
\newcommand{\GTruth}{D\xspace}   
\newcommand{\karticle}{\lambda} 

\newcommand{\fsize}{0.64}
\newcommand{\fsizezwei}{0.46}
\newcommand{\fsizedrei}{0.30}

\newcommand{\bis}{\text{Random Forest}\xspace}

\newcommand{\role}{\text{role}\xspace}

\newcommand{\roles}{\text{roles}\xspace}
\newcommand{\Roles}{\text{Roles}\xspace}

\newcommand{\tsize}{\small}

\begin{document}

\title{TrollSpot: Detecting misbehavior in commenting platforms}


\author{Tai Ching Li}
\affiliation{
  \institution{University of California - Riverside}
  \streetaddress{900 University Ave}
  \city{Riverside} 
  \state{CA} 
  \postcode{92557}
}
\email{tli010@cs.ucr.edu}
\author{Joobin Gharibshah}
\affiliation{
  \institution{University of California - Riverside}
  \streetaddress{900 University Ave}
  \city{Riverside} 
  \state{CA} 
  \postcode{92557}
}
\email{jghar002@cs.ucr.edu}
\author{Evangelos E. Papalexakis}
\affiliation{
  \institution{University of California - Riverside}
  \streetaddress{900 University Ave}
  \city{Riverside} 
  \state{CA} 
  \postcode{92557}
}
\email{epapalex@cs.ucr.edu}
\author{Michalis Faloutsos}
\affiliation{
  \institution{University of California - Riverside}
  \streetaddress{900 University Ave}
  \city{Riverside} 
  \state{CA} 
  \postcode{92557}
}
\email{faloutsos@cs.ucr.edu}

\input{000_abstract}

\maketitle



\input{100_intro}

\input{200_data}

\input{300_metric}

\input{400_identification}
\input{500_behavior}

\input{600_related}

\input{700_conclusion}

\bibliographystyle{ACM-Reference-Format}
\bibliography{main}

\end{document}

%% file: 000_abstract.tex
\begin{abstract}

Commenting platforms, such as Disqus, have emerged as a major online communication platform with millions of users and posts. Their popularity has also attracted parasitic and malicious behaviors, such as trolling and spamming.
There has been relatively little research on modeling and safeguarding these platforms.
As our key contribution, we develop a systematic approach to detect malicious users on commenting platforms focusing on having: (a) interpretable, and (b) fine-grained classification of malice. 
Our work has two key novelties:
(a) we propose two classifications methods, with one following a two stage approach, which first maps observable features  to behaviors 
and then maps these behaviors to user roles,
and (b) we use a comprehensive set of 73 features that span four dimensions of information.
We use 7 million comments during a 9 month period, and we show that our classification methods can distinguish between benign, and malicious roles (spammers, trollers, and fanatics) with a 0.904 AUC.
Our work is a solid  step towards ensuring that commenting platforms are a safe and pleasant medium for the exchange of ideas.

\end{abstract}


%% file: 100_intro.tex
\section{Introduction}\label{sec:intro}

Any successful medium eventually attracts abusive behaviors, and commenting platforms is no exception.
Over the last decade,  commenting on news articles has emerged as a new form of highly social interaction.  First, 
a small number of companies facilitate the backend management of comments for a wide range of websites.
We use the term  {\bf commenting platform} to refer to such platforms, which include Disqus~\cite{disqus}, LiveFyre~\cite{livefyre}, and IntenseDebate~\cite{intensedebate}.
Second, commenting is an intense  activity for many users, who can spend many hours daily at it.


We list a set of definitions that we use in this paper.
A {\bf user} is defined by a platform account, which enables her to leave comments to articles on a website that uses the commenting platform.
A user may leave multiple comments for an article, which leads us to define the {\bf  engagement} of a user for that article.
An engagement has a time duration and intensity in terms of number of comments.
In lack of a better term,
when two \cmms comment on the same article, we say that they {\bf collaborate} and we use the term {\bf collaboration} to describe this activity. 
We use the term  {\bf collaboration intensity}
referring the number of articles for which two users collaborate.

The key question in our work is: {\em Can we automatically detect  malicious users in these commenting platform?}
 Specifically, we frame the problem as follows.
 The input  is the commenting information of the users. This includes: the author of the comment, the time it was posted, and information on the article it was posted for. 
The goal is to identify malicious behaviors and users. 
Detecting parasitic and abusive behaviors is a critical building block for ensuring that these platforms serve their primary purpose,  which is the honest and safe exchange of opinions among readers.




Commenting platforms have attracted little attention so far with only  few exceptions \cite{cheng2015antisocial}\cite{chenganyone}. Most work on modeling and misbehavior detection focuses on Online Social Networks (OSNs), and blogs. In Section~\ref{sec:related}, we survey the key  related areas including: (a)   detecting
 abusive behaviors and malware propagation in OSNs\cite{chu2012detecting}\cite{morstatter2016new}; (b) modeling online user behavior\cite{ferraz2015rsc} \cite{devineni2015if} \cite{gu2016web}\cite{Joobin2017} \cite{Reza2016};
 and (c) analyzing text of online users\cite{sculley2007relaxed}\cite{mishne2005blocking}\cite{sureka2011mining}.  

We propose a systematic comprehensive methodology to identify malicious users on commenting platforms to enable: (a) interpretable, and (b) fine-grained classification of malicious behavior. 
We claim four  key novelties in our work.

{\bf a. A behavior-based  classification.} We propose two classification methods, the first one use typical Random Forest classifier yet another one introduces a two-stage classification approach. In this method, we map: (a) observable features to behaviors, and (b) behaviors into user roles, using unsupervised and supervised learning respectively.

{\bf b. A comprehensive multidimensional feature set.}
We combine 73 features from four different
dimensions of  user interactions: (a) social interaction or user-user interaction, (b) engagement or user-article interaction,  (c) temporal features, and (d) linguistic features.

{\bf c. Fine-grained malicious role identification.}
Our approach goes beyond a  good versus bad determination to a more fined-grained classification
of misbehaving roles.
Here,  we focus on three roles: (a) spammers, (b) trolls, and (c) fanatics, which are defined in the next section. However,  it is easy to introduce more roles as long as appropriate ground truth is available.

{\bf d. Interpretable classification.} The results of our  approach are  interpretable, since they are behavior-centric.
A system administrator can better understand or tweak the definition of, say, a spammer, by looking at the behaviors that constitute its definition (e.g. high number of repeated text across articles, low engagement per article).

Our study is grounded on nearly 7 million comments from nearly 200K users over 9 months from Discus, which is arguably the largest commenting platform. Here we highlight some interesting result from our study.



\textbf{a. Identifying misbehaving users with 0.904 AUC.} Our classifier can efficiently identify misbehaving users and it outperforms the baseline we compare with. 

\textbf{b. The \role classification result achieves 80.8\% overall accuracy.} 

c. \textbf{We find patterns of misbehaving users from their interpretable \lub.}  An example \lub indicates that making longer comments and using capital letters more frequently is a sign of misbehaving users, and the later the active hour of a users, the more likely they are to misbehave.

\textbf{d. Large scale study.} About 0.9\% of total users in our data-set are misbehaving users, and CNBC has the largest percentage of misbehaving users which is 1.48\%. Only 15\% of the misbehaving users exhibit a cross-website behavior.

%% file: 200_data.tex
\section{Data collection and definitions}\label{sec:dataset}
Disqus is one of the most widely-used commenting service provider today with  a billion unique visitors a month and installed by more than two and a half million sites, including ABC News, Rolling Stone, IGN, Bloomberg and more. 


\textbf{Data Collection.} We collected  data from Disqus  through its  Application Programming Interface (API). 
The API can be used to collect all the comments for a given article, and for all the articles of a given website. However, the functionality of collecting 
 all the comments for a given user would have been useful for our study, but it became unavailable in 2014 for privacy reasons.

Using the website-centric API, 
we collect data from four popular  websites: 
(a) CNBC News,
(b) ABC News,
(c) Bloomberg Views
  and (d) Breaking News - a Disqus channel.
The first three are well-known news websites and the last one is the most popular channel on Disqus. A channel is similar to a newsfeed, whose  articles  are selected by the users that participate in that channel.
We collect all comments posted at articles published on these 4 sources in between Nov 1st 2015 and July 31st 2016. The dataset consists of: (a) 286,275 articles, (b) 6,994,693 comments and (c) 201,112 unique users, (d) 1,705,667 engagements.



\textbf{Establishing the ground truth of misbehaving users.} 
A key challenge in our work is the need for ground truth of misbehaving users. 
Fortunately, Disqus enables users to report misbehaving comments and the  number of reports a comment received is provided by Disqus API. 
In our 7M comments, we have 66.4k comments reported as ``bad/inappropriate" by the community. We will further discuss this in Section~\ref{sec:clustering} and show how we leverage this to identify misbehaving users.

\textbf{\Roles of misbehaving users and ground truth.} We identify and attempt to detect three different \roles of misbehaving users: trolls, spammers and fanatics.
These are inherently difficult to define, so we resort to human feedback.

We define three malicious roles below. We show here the definitions that we gave our evaluators, as we explain in Section~\ref{sec:clustering}. 
\begin{enumerate}
\item \textbf{Trolls.} Users who make inflammatory or inappropriate comments for the sole purpose of upsetting other users and provoking a response.
\item \textbf{Spammers.} Users who repeatedly make similar comments in the same or multiple articles.
\item \textbf{Fanatics.} Users who exhibits an extreme and uncritical enthusiasm in religion or politics.
\end{enumerate}

%% file: 300_metric.tex
\section{Features and user behavior}\label{sec:metric}

In this section, we identify and study features that relate to user behavior on the Disqus platform. Although studying each of these features in depth is of independent interest, the goal is to identify meaningful features that can help us detect misbehavior. 
Due to space limitations, we can only provide highlights of the features and their distributions.
In Table~\ref{tab:features},  we outline the features that we use in 
Section \ref{sec:clustering}.

We study the behavior of users along four dimensions: \textbf{(a)} engagement behavior (user-article interaction), \textbf{(b)} social behavior (user-user interaction), \textbf{(c)} temporal behavior, and \textbf{(d)} linguistic properties. 

\begin{table*}[!t]
	\tsize
    \renewcommand{\arraystretch}{1.3}
    \centering
    \begin{threeparttable}[b]
      \caption{The overview of the 73 features we use per Dimension}
      \label{tab:features}      
      \begin{tabular}{p{2cm}|p{11.5cm}|p{1cm}}
          \hline
          Dimension & Features & Count \\
          \hline
          Engagement & number of engagements, engagement duration\tnote{*}, engagement intensity\tnote{*} & 7 \\
          Social & degrees, number of maximal cliques, number of triangles (in different level of collaboration intensity)  & 17 \\ 
          Temporal & number of comments made in 24-hour slots, \rh & 25 \\
          Linguistic & number of words\tnote{*}, number of sentences\tnote{*}, percentage of capital letters\tnote{*}, readability metrics, number of URLs & 24 \\
          \hline
      \end{tabular}
      \begin{tablenotes}
      	  \item[*] We use several statistical versions (mean, maximum and minimum)  of the feature per engagements, user or comment. 
      \end{tablenotes}
  \end{threeparttable}
\end{table*}

\input{301_engagement}

\input{302_social}
\input{303_temporal}
\input{304_linguistic}

%% file: 301_engagement.tex
\subsection{Engagement behavior}

\begin{figure}[ht]
	\centering
	\includegraphics[width=\fsize\columnwidth]{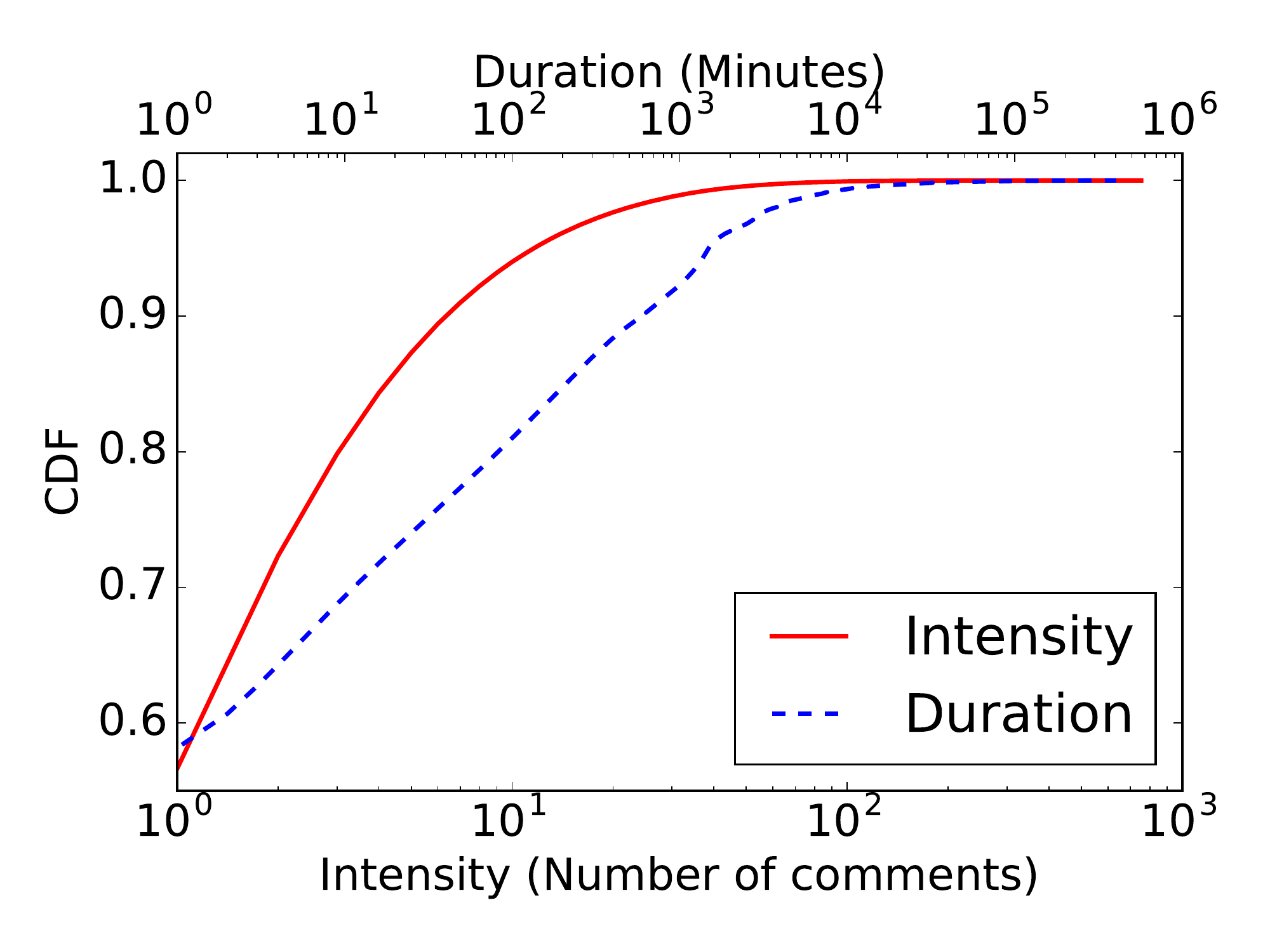}
    \caption{\textbf{Engagement behavior:} Distribution of intensity and duration of engagements.}
    \label{fig:engagement_intensity_dureation}
\end{figure}

We quantify the engagement (user - article interaction) with 7 different features. Six of them are derived from two major properties: The \textbf{engagement duration} is the time interval between the first comment and the last comment user makes on the article. 
If the user leaves only one comment, we consider this as zero length interval.  
The \textbf{engagement intensity} is the total number of comments user makes on the article.

\textbf{Engagement duration: 90\% last for less than 10 hours, but some can be as long as half an year.}
In Figure~\ref{fig:engagement_intensity_dureation}, we plot the Cumulative Distribution Function (CDF) of the  duration (top x-axis) for all 1.7M engagements. We find that 90\% of engagements last less than 10 hours and have less than 7 comments. Interestingly, we find  106 engagements which last for more than half year!

{\bf Engagement intensity: 90\% have less than 7 comments, but 0.06\% have more than 100 comments.} 
In Figure~\ref{fig:engagement_intensity_dureation}, we plot the CDF of the intensity (bottom x-axis) for  all engagements. We find that 90\% of them have less than 7 comments, while 1,151 (0.06\%) of them have more than 100 comments.

\textbf{High intensity is correlated with misbehavior.} The observations above raise the question: is unusually high engagement intensity associated with misbehavior? Preliminary manual inspection suggests that these engagements contain many reported (by the community)  comments. We quantify the correlation between engagement intensity and the total number of reported comments in the engagement and we find a Pearson correlation coefficient  $\rho=0.53$. 
This suggests that engagement intensity should be a good metric in our classification.

%% file: 302_social.tex
\subsection{Social Behavior}


We propose 17 features to model the social interaction of the users, which we define as the posting of comments to the same articles.

{\bf Single-article collaboration Threshold: $\theta$ comments.} We say that two users collaborate in one article, if they each post at least $\theta$ comments on that article. For $\theta =1$, the graph become very dense, and the analysis is both slow and less informative. In the remaining of this work, we use a threshold  $\theta =2$.

{\bf User-user collaboration intensity and threshold: $\karticle$ articles.} 
The collaboration intensity is the number of articles that two users collaborate for a given threshold $\theta$. To study collaborations at different levels of intensity, we introduce the {\bf collaboration intensity threshold $\karticle$}, which we use below.

We define  the undirected weighted {\bf collaboration graph} $G_\karticle = \langle V_\karticle, E_\karticle \rangle$  of collaboration intensity $\karticle$ where:
\begin{enumerate}
	\item $V_\karticle$ is a set nodes $v$, representing users.
	\item $E_\karticle$ is the set of edges, where  edge $e_{ij}$ between nodes $v_i$ and $v_j$ exists,
		  if and only if the collaboration intensity of the users exceeds the  {\bf threshold of  $\karticle$ articles}. 
          The edge weight  $w(e_{ij})$ is set to  the collaboration intensity.
\end{enumerate}

The collaboration graph (for $\theta=2$ and $\karticle = 0$) has 95,527 users and roughly 21 millions edges, an average degree of 440.7 and a median degree of 137. Note that we do not include users with zero degree  in this graph.

\begin{figure}[!tb]
	\centering
    \subfigure[]{
    	\includegraphics[width=\fsizezwei\columnwidth]{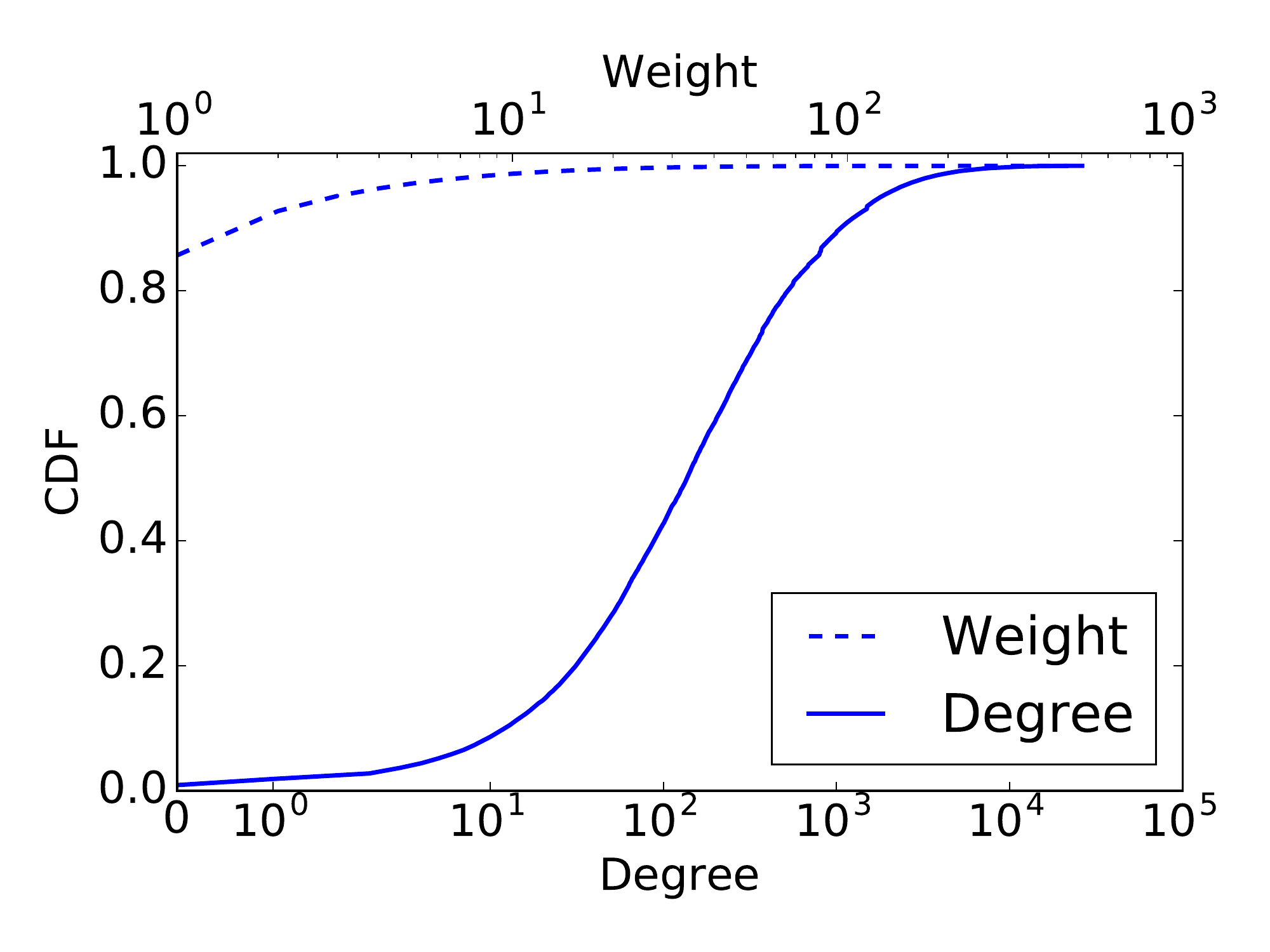}
        \label{subfig:social_degree_weight}
    }
    \subfigure[]{
    	\includegraphics[width=\fsizezwei\columnwidth]{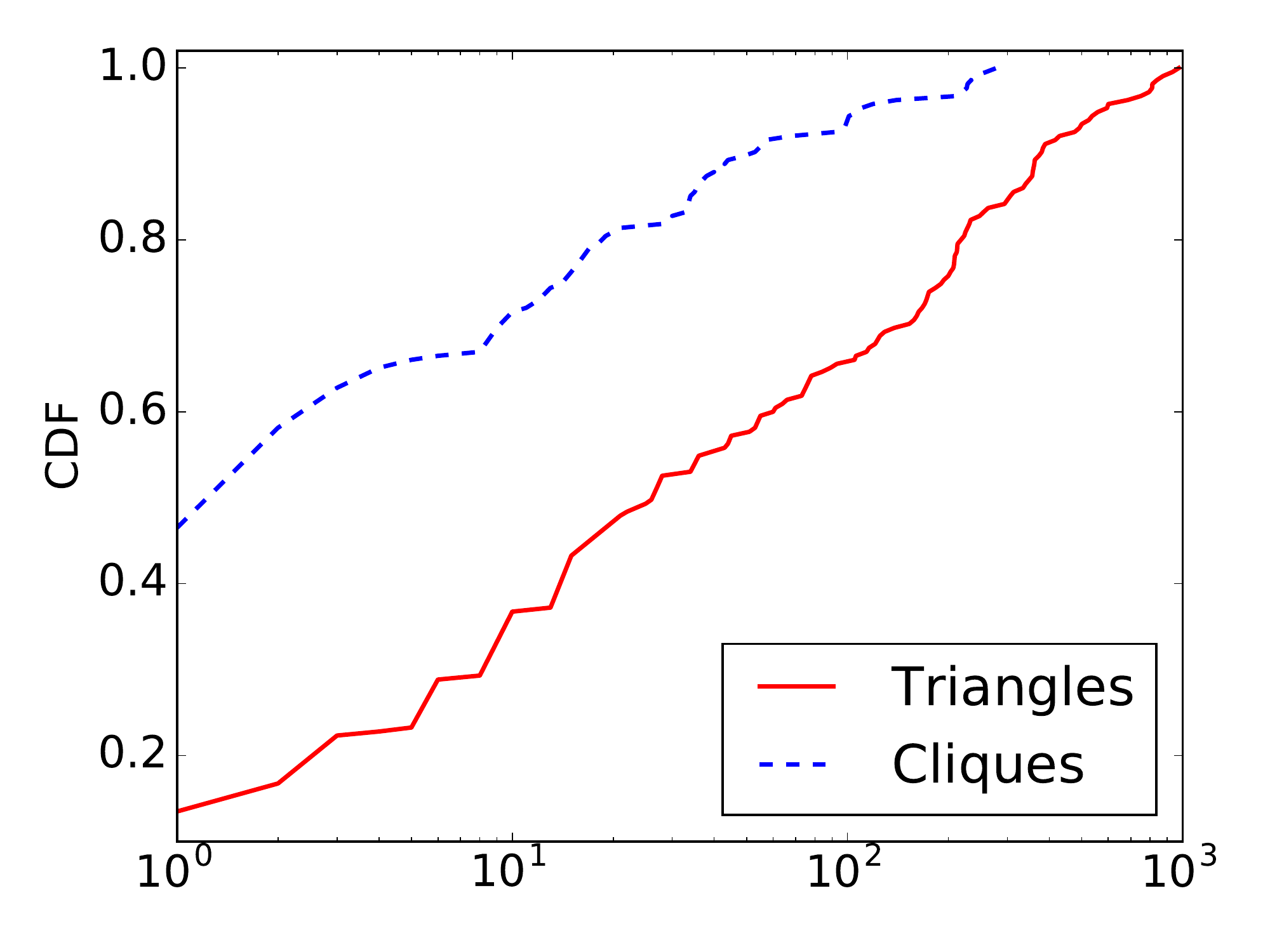}
        \label{subfig:social_local_density}
    }
   	\caption{\textbf{Social behavior:} (a) the CDF of the user degrees (bottom x-axis) and the edge weight distributions (top x-axis), 
    (b) CDF of the number of cliques and triangles of user in the $G_{128}$ collaboration graph.}
\end{figure}

In Figure~\ref{subfig:social_degree_weight}, we plot the CDF of the user degrees (bottom x-axis) and the edge weight distributions (top x-axis). We see that 90\% of the users have degree lower than 1,054, the max degree goes to 26,318: this user collaborates with more than 27\% of users in the graph! The figure also shows that 90\% of the edges have weights less than 2. Although it is not easy to gauge from the plot, we find 12,374 edges with weight over 64, 1,779 edges with weight over 128 and 65 edges with weight over 256. In other words, there are 65 pairs of users who have collaborated on   more than 256 articles. 


\textbf{Capturing the collaboration groups: triangles and cliques.}
We quantify the local connectivity of the users using the number of maximal cliques and triangles in which a user participates. Both these metrics capture how densely connected the neighbors of that user are.  Figure~\ref{subfig:social_local_density} shows the distribution of triangles and maximal cliques of user in graph, for $\karticle$ equals to 128.  We do not count "trivial" cliques of size two. We see that 50\% of users have less than 27 triangles on the neighborhood, 90\% of have less than 376, but there are 1\% with more than 868 triangles. 
We also see that 90\% of users have less than 51 maximal cliques and there are 8 users, who participate in more than 186 cliques which is more than 50\% of all maximal cliques in the graph.
These  highly collaborative users are suspicious and this encouraged us to consider both these features for misbehavior detection.

%% file: 303_temporal.tex
\subsection{Temporal behavior}

We quantify the temporal behavior of users with 25 features.

\begin{figure}[!b]
	\centering
    \subfigure[]{
    	\includegraphics[width=\fsizezwei\columnwidth]{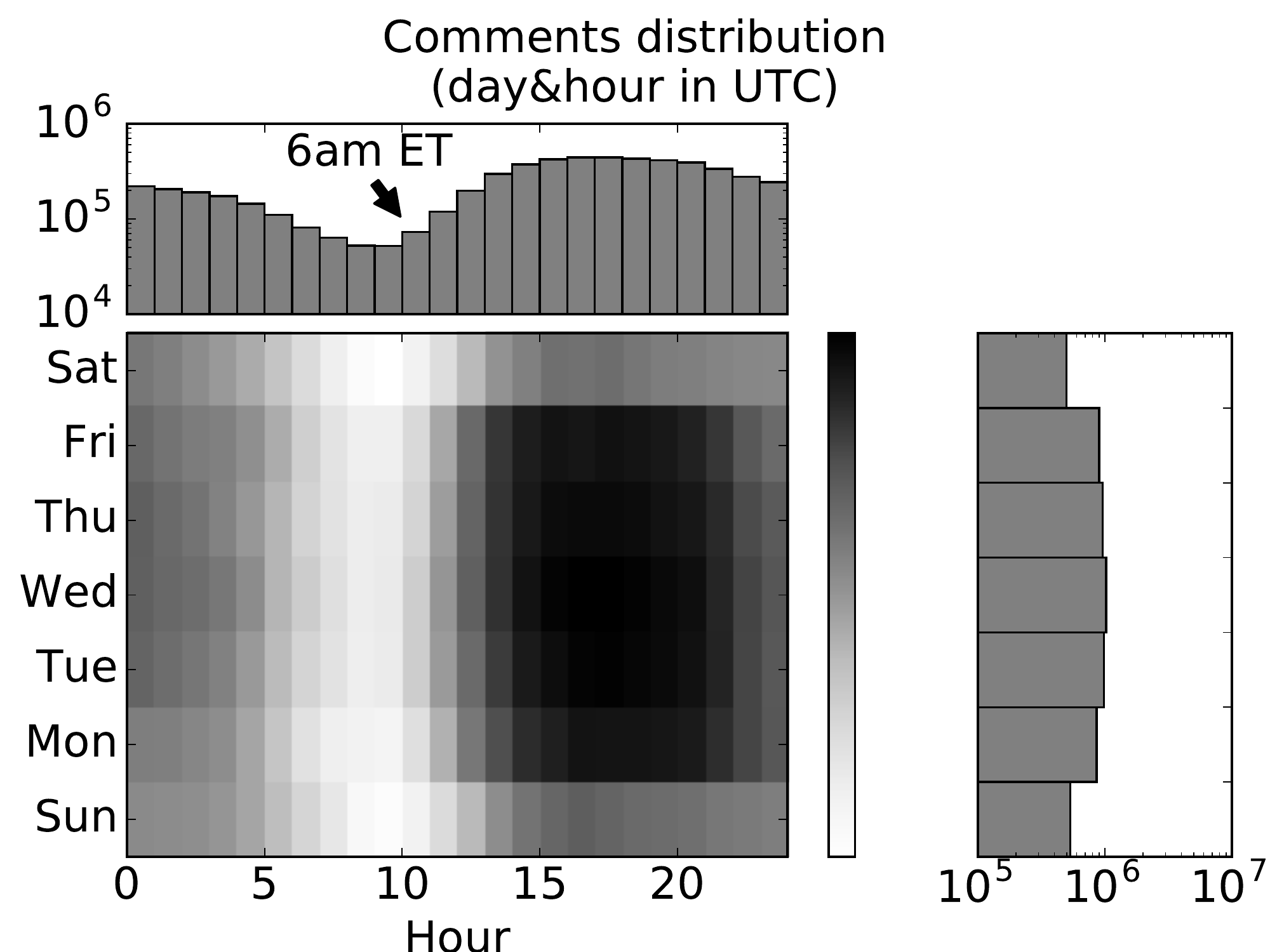}
        \label{fig:periodicity_heatmap}
    }
    \subfigure[]{
    	\includegraphics[width=\fsizezwei\columnwidth]{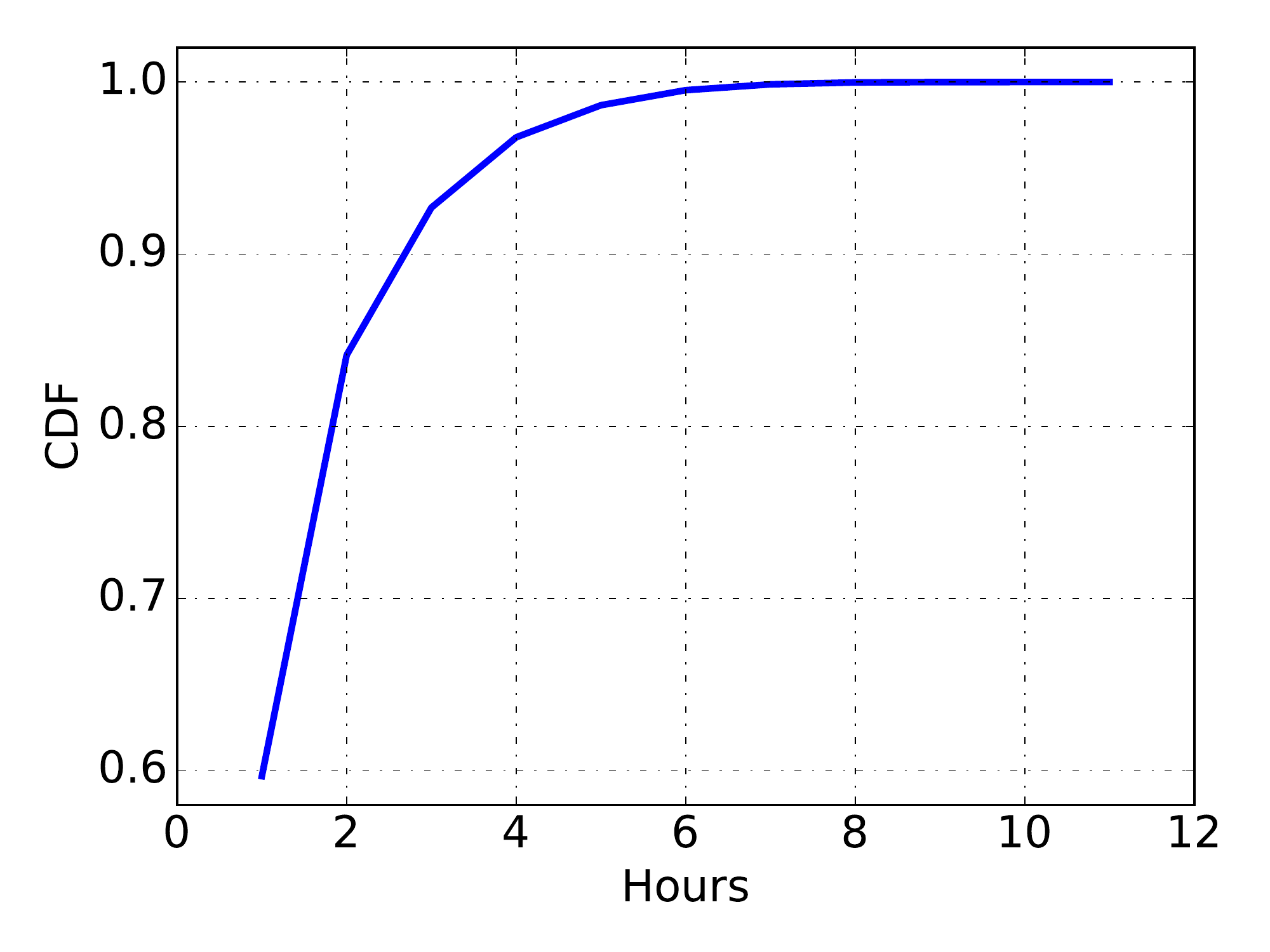}
        \label{fig:rush_hour}
    }
   	\caption{\textbf{Temporal behavior.} (a) Time and day of the week plot. (b) The distribution of \rhs of users.}
\end{figure}

\textbf{Most users exhibit persistent  behavior: daily and weekly.} 
We use the {\bf time and day of the week} plot to understand the temporal behavior of the users. The plot shows the number of comments in each hour-of-day and day-of-week as a heat-map. Figure \ref{fig:periodicity_heatmap} shows this behavior for all the users, but individual users exhibit similar  daily and weekly behavior, as we will see below. On the x-axis (hour-of-day), we see that the commenting activities increase at the beginning of day in the east coast as annotated in the plot, the peak hours cross the noon period of both east and west coasts then drop sharply after that. Given the news websites we are studying, it is reasonable to assume a US-centric user majority. In y-axis (day-of-week), the activity increases at the start of the week, peaks in the middle, and drops before the weekend. Combining the two observations,  we see that Disqus users post most of their comments during the common work-hours in the week. Note that the same pattern is also observed on other social networks, like Facebook~\cite{warren2010fb} and Twitter~\cite{sysomos2014tw}.

The plot lead us to the following hypothesis:

{\bf Hypothesis:} A typical user makes most of her comments in a period of 3-4 hours in a day. 

\textbf{The \rhs of users commenting behavior.} To assess the validity of our hypothesis, we would like to answer the question: how many hours a day does a user spend commenting? Determining this  hides subtleties as: (a) patterns can vary from day to day, and (b) it could be that many hours have non-zero activity due to averaging over a long period.
Thus, we opt to capture how ``focused" or "spread" is the commenting activity of the user. We define {\bf \rhs } to be
 the minimum number of hours, during which the  user makes more than 50\% of their total comments during a day on average. 

{\bf \Rhs is less than 4 hours for 96.7\% of the users.}
The distribution of the users' \rhs is shown in Figure \ref{fig:rush_hour}. 
The plot shows that 96.7\% of users have  \rhs  less than four hours, which matches our hypothesis. Interestingly, we find 368 users who have more than 8 \rhs,  a significantly wider spread. Upon inspection, many of these users have non-trivial activity over 14 hours in a day. This wide range of behaviors suggests that \rhs can be a useful metric for our classification.

%% file: 304_linguistic.tex
\subsection{Linguistic properties}\label{subsec:linguistic}

\begin{figure}[!htb]
	\centering
    	\includegraphics[width=\fsize\columnwidth]{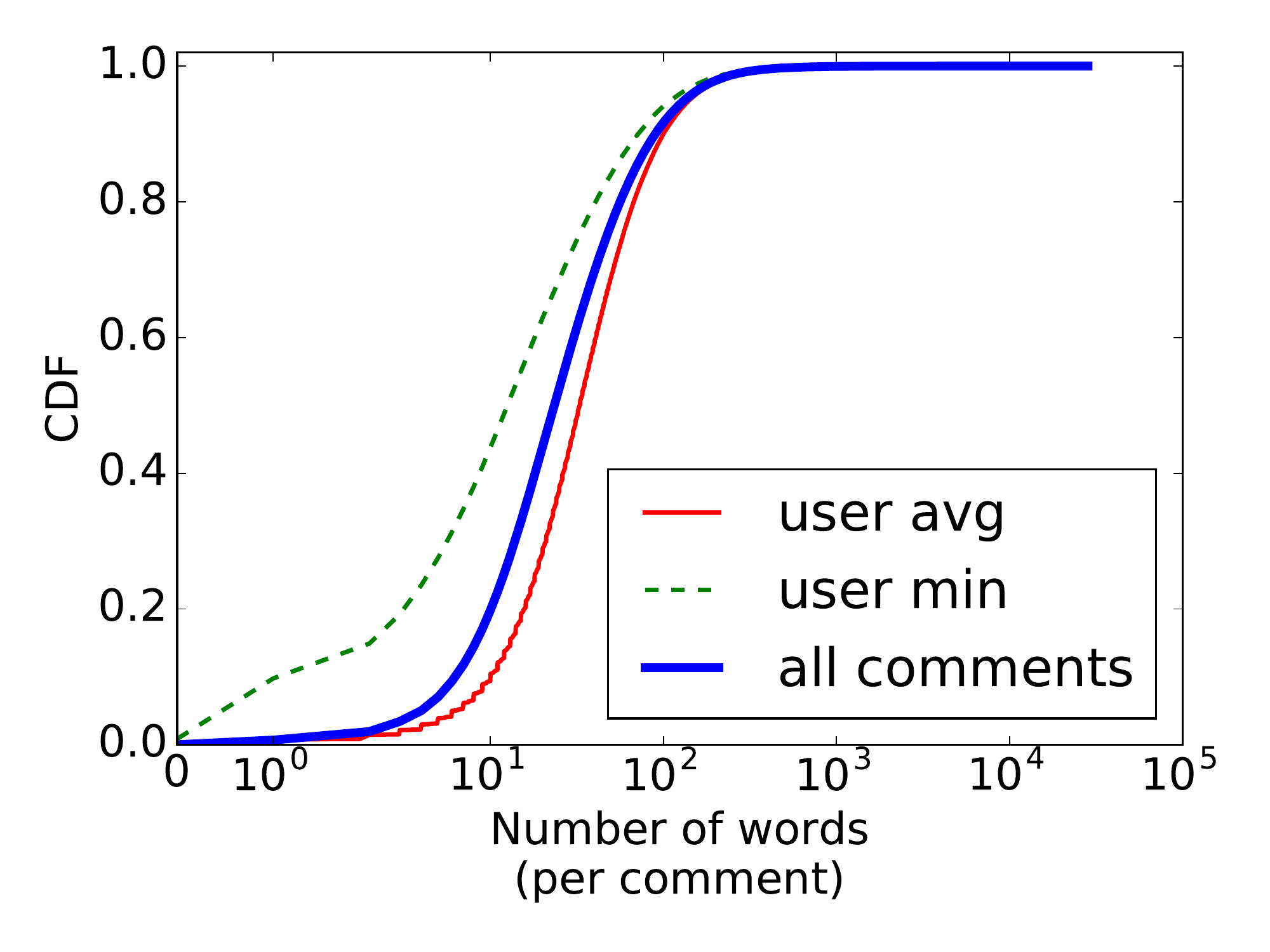}
        	\caption{\textbf{Linguistic property}: number of words in a comment}
    \label{fig:length_comment}
\end{figure}


We identify and study 24  linguistic features from the text of the posts, such as the length of the comment, several metrics readability, number of URLs, but discuss only a two metrics.

{\bf Only 1.9\% of comments contain URLs.} 
We naturally consider the existence of  URLs in a comment as a feature in our classification: their presence can indicate  ad-oriented spamming. 
We find that only 1.9\% of all comments  contain one or more URLs.
We also find that only  1.7\% of the users have ever posted more than 3 comments containing URLs.
In fact, our interaction with the data suggests that often users will use URLs as references in support of their opinions. 

\textbf{Length: 91\% of comments have less than 100 words.} The length of a  comment is a good metric for the interest and the time a user is willing to spend expressing her opinion about the article.  In Figure~\ref{fig:length_comment}, we plot the distribution of the number of words  in a comment in three different ways: \textbf{(a)} across all comments, \textbf{(b)} the average comment length per user,  and \textbf{(c)}  minimum comment length per user.
We see that 91\% of comments have less than  100 words, while
roughly 20\% of the comments with less than 10 words.
Interestingly, we also find 14,838 (0.002\%) comments with more than 500 words, which is roughly equivalent to 1.5 pages (per wordstopages.com with Times New Roman, 12 font-size, single-spacing): the comment is the size of small article! 

\textbf{Lengthy comments are more likely to be spam.} We examine the lengthy comments (>500 words) and find out that 47\% of them are verbatim copies of at least one other comment from the same user in our dataset. This is an indication of spamming especially in its broader definition that we adopt here.  Such phenomenon is  more pronounced for higher word counts: 63\% of comments have verbatim copies for length over 1,000 words , and 83\%  of comments with length  2,000 words or longer. This suggests that comment length is a helpful feature in detecting misbehavior.

%% file: 400_identification.tex
\section{Feature-based Misbehavior Identification}\label{sec:clustering}

Having identified interesting features,  we develop a method to   identify misbehaving users and their effectiveness. For  convenience, we outline the features in Table~\ref{tab:features}, and we intend to define them in detail in an extended version of this paper.

\input{402_methodology}
\input{403_roles}

\input{404_study}

%% file: 402_methodology.tex

\begin{figure}[!htb]
	\centering
    \subfigure[]{
    	\includegraphics[width=\fsizezwei\columnwidth]{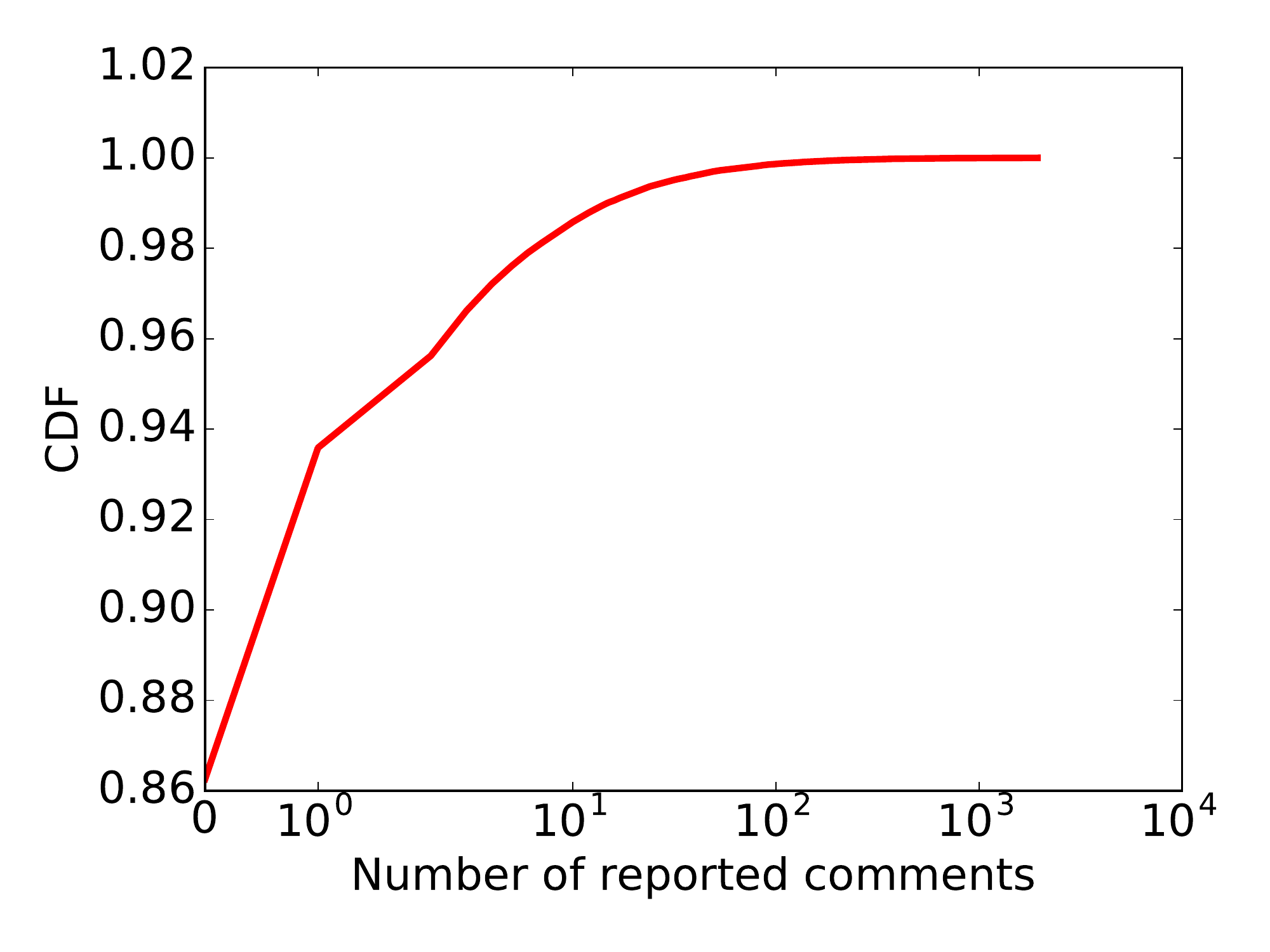}
        \label{fig:reported_users}
    }
    \subfigure[]{
    	\includegraphics[width=\fsizezwei\columnwidth]{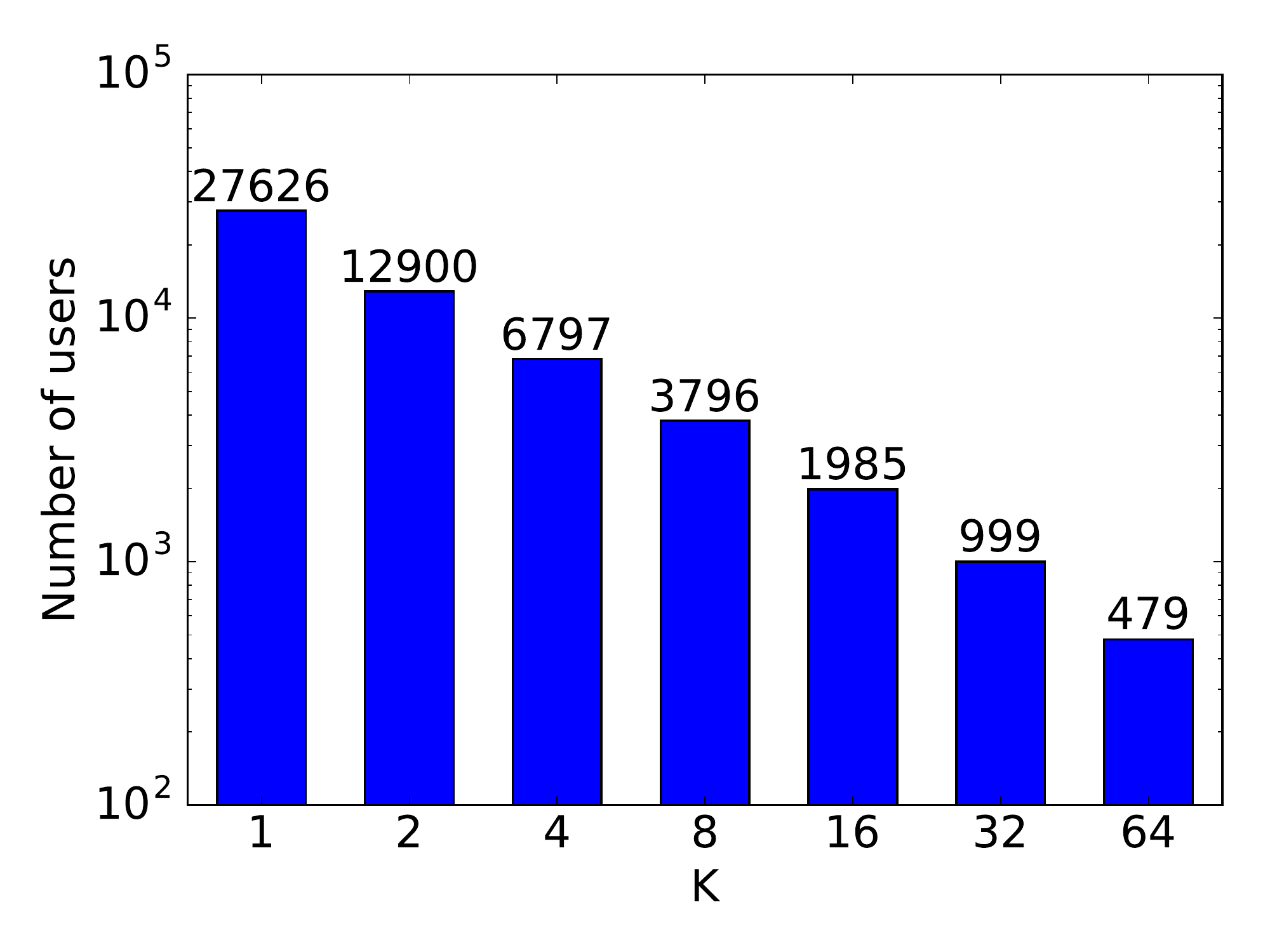}
        \label{fig:reported_users_with_k}
    }    
   	\caption{(a) Distribution of  \rcmms of a user. (b) Number of users having $k$ \rcmms .}
\end{figure}

\textbf{A. Establishing the ground truth.} 
To overcome the lack of absolute truth,  we rely on ``proxy signals". 
Here, we use the community's own opinion: any user can report (a.k.a. flag) a comment as ``inappropriate". Below, we explain how we use this community feedback to construct the ground truth, as the process hides several subtleties.

{\bf Ground-truth: Reportings per malicious comment.} 
To increase our confidence, we set a minimum threshold  of reports, $\nrep$, that a comment must have to labeled malicious.
The rationale is that a single reporting can be created even accidentally (the authors have regrettably done this once).
After analysis and deliberation omitted here, we settled on $\nrep = 3$ reportings. 
We use the term {\bf reported comment} to refer to a comment with more than $\nrep$ reports.

{\bf Ground-truth: Reported comments per malicious user.} In the same vain, we want to be careful in labeling a user as malicious based on the number of reported comments.
We use the {\bf \ncommThr}, $\ncomm$, to control tune the definition of malicious user.


\textbf{Roughly 86\% of the users have no \rcmms.} 
We plot the distribution of the total number of \rcmms for each user  in Figure~\ref{fig:reported_users}. We  find that 86.2\% of users have no \rcmms,    and only 1.6\% of users have more than 10 \rcmms. 

We consider users with zero \rcmms as {\bf benign} for the purpose of establishing the ground truth.

{\bf Building the ground truth datasets: $\GTruth_\ncomm$.}
We create a set of labeled datasets as follows.
First, we distinguish \rusers into groups, $\VReported_{\ncomm(i)}$, 
with $\ncomm(i) = 2^i, for ~~ i = 0, 1, ...$
A user is in group $\VReported_{\ncomm(i)}$,  if the number of her \rcmms are greater or equal $\ncomm(i)$. 
It turns out that
 no user has more than 128 reported comments. The numbers of users in each group with threshold $\ncomm(i)$ are shown in Figure~\ref{fig:reported_users_with_k}. 
Second, we create datasets $\GTruth_{\ncomm(i)}$ by randomly selecting 200 \rusers from $\VReported_{\ncomm(i)}$, and combining them with 200  benign users (zero \rcmm). 

\textbf{Benign user selection: Minimizing the effect of the number of comments of a user.}
When selecting the 200 benign users to build the datasets,
 we follow the insightful approach of the earlier in this space~\cite{cheng2015antisocial}.
Namely, we find benign users that ``match" the number of comments of \rusers. The goal is to minimize the dominant role that the number of comments of a user can play.
First, the number of comments and number of \rcmms of a user are strongly positively correlated with a
 Pearson Correlation Coefficient of 0.73.
 Second, most users have few comments given the skewness of the distribution.
 Thus, a purely random selection of benign users would have probably created a low-activity benign set in terms of user comments, and a highly-active \rusers.
 The classification would have been very accurate by relying heavily  on the number posts.


\begin{figure}[!htb]
	\centering
    \subfigure[]{
    	\includegraphics[width=\fsizezwei\columnwidth]{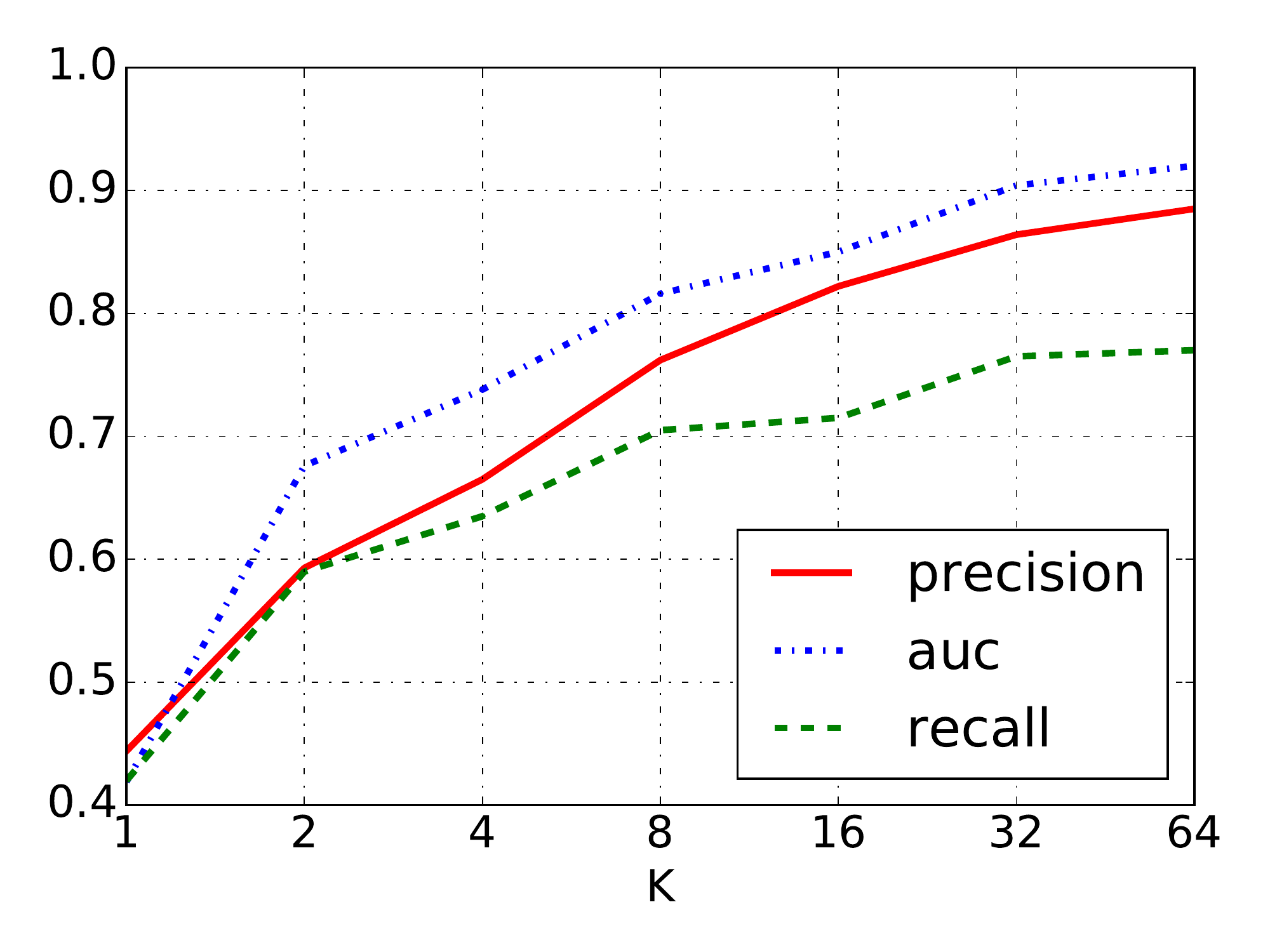}
        \label{fig:all_result}
    }
    \subfigure[]{
    	\includegraphics[width=\fsizezwei\columnwidth]{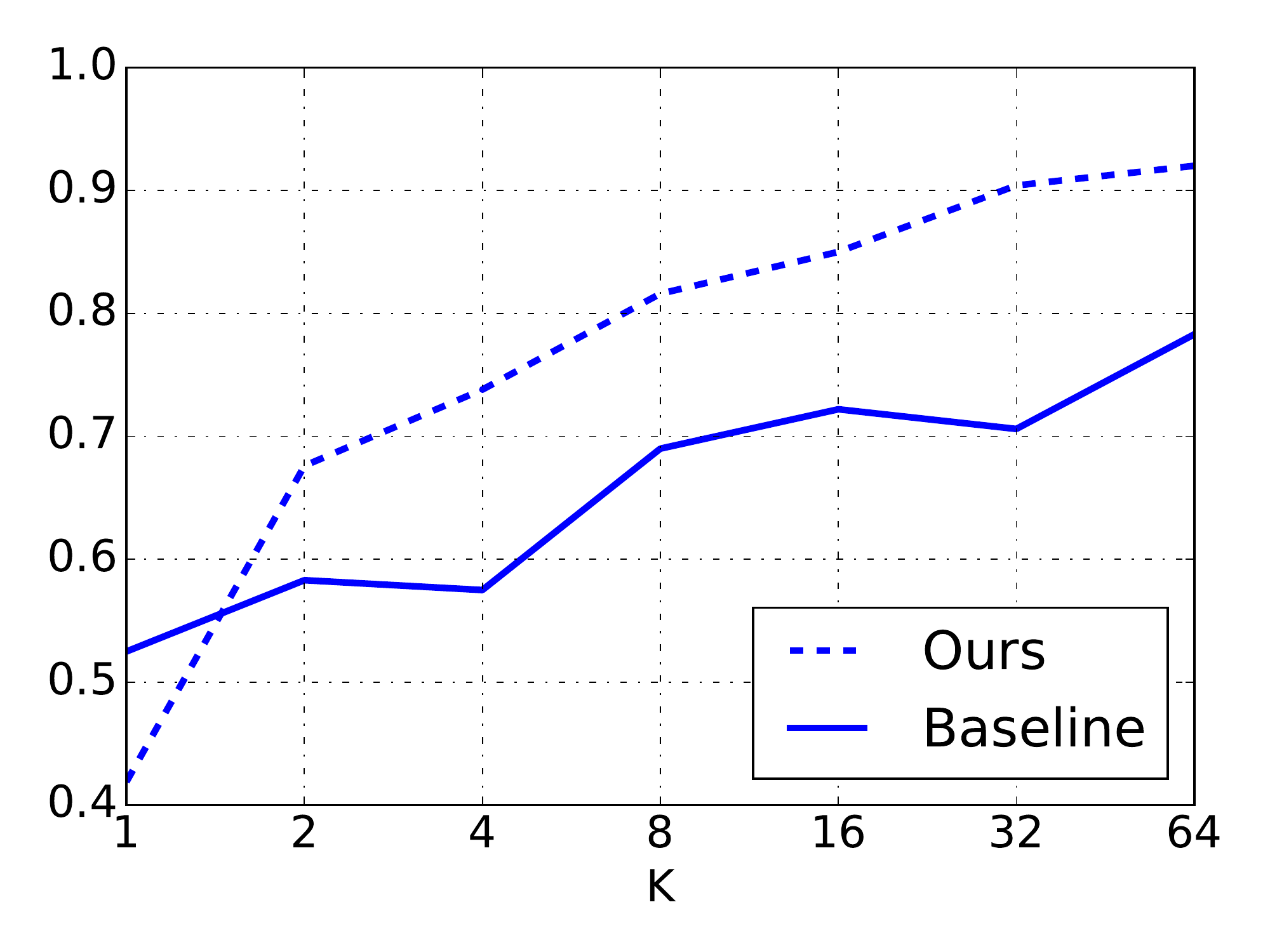}
        \label{fig:all_vs_jurelite}
    }
   	\caption{(a) Classification results as a function of the number of reported comments that ``incriminate" a user. (b) AUC of our approach and the baseline.}
\end{figure}

\vspace{.2cm}
{\bf B. The benign-malicious classification.}
For the classification, we use the \bis classifier provided by Weka \cite{witten1999weka}, which gave the best results among many that we tried. 
We perform ten-fold cross validation and report the precision, recall, and {\bf ROC curve (AUC)} of each dataset in Figure~\ref{fig:all_result}. The plot shows that our features can identify \rusers with more than 80\% precision when the threshold $\ncomm$ is larger than 16. 

{\bf Selecting \ncommThr $\ncomm = 32$}.
We manually examined \rusers in $\GTruth_{16}$, $\GTruth_{32}$ and $\GTruth_{64}$ by sampling 20 users from each group. 
We find that users with more than 16 \rcmms exhibit a persistent misbehavior throughout their lifetime. The 40 \rusers sampled from $\GTruth_{32}$ and $\GTruth_{64}$ are 100\% labeled as misbehaving users by our independent human evaluators.  This gives us confidence to claim that users with  32  or more \rcmms are misbehaving users. Thus, we  use  dataset $\GTruth_{32}$ as reference below.

\textbf{The baseline classifier.} 
We  adapt and use a previously proposed algorithm as a reference in our study \cite{cheng2015antisocial}. That work focuses on predicting whether a user will get banned in the future, which is a somewhat different goal from our work.  

We  faithfully reproduce the reference method  \cite{cheng2015antisocial}, which also uses 
 the Random Forest technique like we do.
In their work, they use 35 features, however 6 are not publicly available, namely the website moderator features, such as the number of up-votes given to others (similar to a Facebook ``Like''). Unfortunately, we were not able to access those features through the public API, therefore we do not use them in our experiments.

{\bf The accuracy of our method exhibits 90\% AUC.} 
For clarity of presentation, we only show the comparison of AUC of the two methods in Figure~\ref{fig:all_vs_jurelite}. 
Our method performs better than the baseline classifier, with the caveat that both approaches are restricted to the publicly available features.
The most interesting conclusion is that our features  have good discriminatory power in classifying misbehaving and benign users.



%% file: 403_roles.tex

\vspace{0.2cm}
{\bf C. The fine-grained malicious role classification.}
The novelty of the work relies partly on going beyond the ``malicious" label to the role of the misbehaving user.

{\bf Ground truth for the roles of misbehaving users.} Here, we resort to manual labeling to create our reference data: we examine 200 misbehaving users in $\GTruth_{32}$ and categorize them into three different roles: trolls, spammers and fanatics. To the best of our knowledge, this is the first work that does such a fine-grained classification of user behaviors, especially in commenting platforms. 

\begin{table}[t]
	\tsize
    \renewcommand{\arraystretch}{1.3}
    \centering
    \caption{Role classification result}
    \label{tab:role_classification}      
	\begin{tabular}{p{2.5cm}|p{2cm}|p{2cm}}
    \hline
    \textbf{Role} & \textbf{Precision} & \textbf{Recall} \\
    \hline
    Trolls & 86.4\% & 73.1\% \\
    Spammers & 58.6\% & 81\% \\
    Fanatics & 86.1\% & 73.3\% \\
    Benign & 81\% & 87.5\%\\
    \hline
    \textbf{Overall accuracy} & \multicolumn{2} {l} {80.8\%}  \\
    \hline    
	\end{tabular}
\end{table}

Each user is labeled by three evaluators who were presented with the definitions in Section~\ref{sec:dataset}. 
In absence of unanimity, we set their role by taking the majority label. 
Out of 200 misbehaving users, 104 are labeled as trolls, 21 as spammers and 75 as fanatics.

{\bf Role classification: Our approach has 80.8\% overall accuracy.}
Having ground truth, we apply 10-fold cross validation with the same classifier and the same 73 features that we used to identify misbehaving users. 
Our result shows that our model can effectively classify the \role of misbehaving users with an overall accuracy of 80.8\% as shown in
Table~\ref{tab:role_classification}.
For all the classes the recall is above 73\%
and the precision above 81\% except the Spammers.



\textbf{The community shows tolerance to non-provocative spammers.} Intrigued by the low precision for spammers, we find that spam comments without provocative language, swear words and sarcasm often do not get reported by the community.
 Recalling Section~\ref{subsec:linguistic},  the unusually long comments are more likely to be spam, due to verbatim repetitions.
However, this behaviors seems to either escape detection or be met with tolerance.

{\bf Our method identifies un-reported spammers.}  We examine the 12 false positive in the spam category: our spam label is not corroborated by the community. We actually find that at least one of these users exhibits clear spamming behavior, as she repeats the exact same comment 3 times in one article and 5 times in another. 
This investigation suggests that our approach could be catching parasitic behaviors that the community could miss. As a result, the accuracy of our algorithm could be better than reported here, especially for spammers.


%% file: 404_study.tex
\vspace{0.2cm}
{\bf D. Studying the misbehaving users in the wild.}

We apply the classier to the whole dataset to understand how misbehaving users distributed in the wild. For identifying misbehaving users, the classifier labeled 1,738 (0.86\%) users as misbehaving with confidence over 80\%. The role classification results are: 866 users are labeled as trolls, 165 users are labeled as spammers and 597 users are labeled as fanatics. We also study how misbehaving users distributed in each websites, the result is shown in Table~\ref{tab:misbehaving_users_websites}. It shows that CNBC has the largest number of misbehaving users with 1,167 which is 1.48\% of total users active on the website while the rest of websites are at most 0.56\% of users are misbehaving users. Again, we only label misbehaving users with confidence over 80\%.

\begin{table}[htb]
	\tsize
    \renewcommand{\arraystretch}{1.3}
    \centering
    \caption{Misbehaving users in websites}
    \label{tab:misbehaving_users_websites}      
	\begin{tabular}{m{2.5cm}|m{2cm}|m{3cm}}
    \hline
    \textbf{Website} & \textbf{Total users} & \textbf{Misbehaving users} \\
    \hline
    ABC News & 129,053 & 717 (0.56\%) \\
    CNBC & 78,618 & 1167 (1.48\%) \\
    Bloomberg Views & 33,223 & 131 (0.39\%) \\
    Breaking-News (Disqus channel) & 11,607 & 48 (0.41\%) \\
    \hline    
	\end{tabular}
\end{table}


%% file: 500_behavior.tex
\section{Interpretable two-stage classification}\label{sec:taxonomy}


In this section, we introduce a novel, two-stage approach in classifying misbehaving users. In the first step we identify tightly-knit groups of users who exhibit very similar behavior across a certain subset of the features we discussed in the previous section. Each such group of users and their associated features defines a {\em latent user behavior}. Note that we do not define those latent user behaviors a-priori, they rather emerge from the data automatically. Subsequently, we use these latent user behaviors as our new features towards classifying misbehaving users. The advantage of this two-stage approach is that the latent user behaviors offer interpret-ability of the results and allow for in-depth analysis of different user (mis-)behaviors.



\input{501_coclustering}
\input{502_methodology}

%% file: 501_coclustering.tex
\subsection{Identifying latent behaviors:  Co-clustering}
The first stage of our two-stage approach can be exactly mapped to an instance of {\em co-clustering}.
In a nutshell, co-clustering is the simultaneous clustering of all data modalities. In our case, consider a representation of a user in the vector space defined by the entire set of their features. Hence, the modalities of our data are two: ``user'' and ``feature''. Co-clustering of a users $\times$ features data matrix essentially yields different groups (or co-clusters), where each one contains a subset of the users and a subset of the features. In other words, when grouping users and features together, co-clustering allows for the flexibility of finding a set of users that are highly similar {\em only for a subset} of their features. This subset of features is key to our method. Essentially, the subset of features of a co-cluster defines a {\bf latent user behavior}. 

\begin{figure}[ht]
\begin{center}
	\includegraphics[width=\columnwidth]{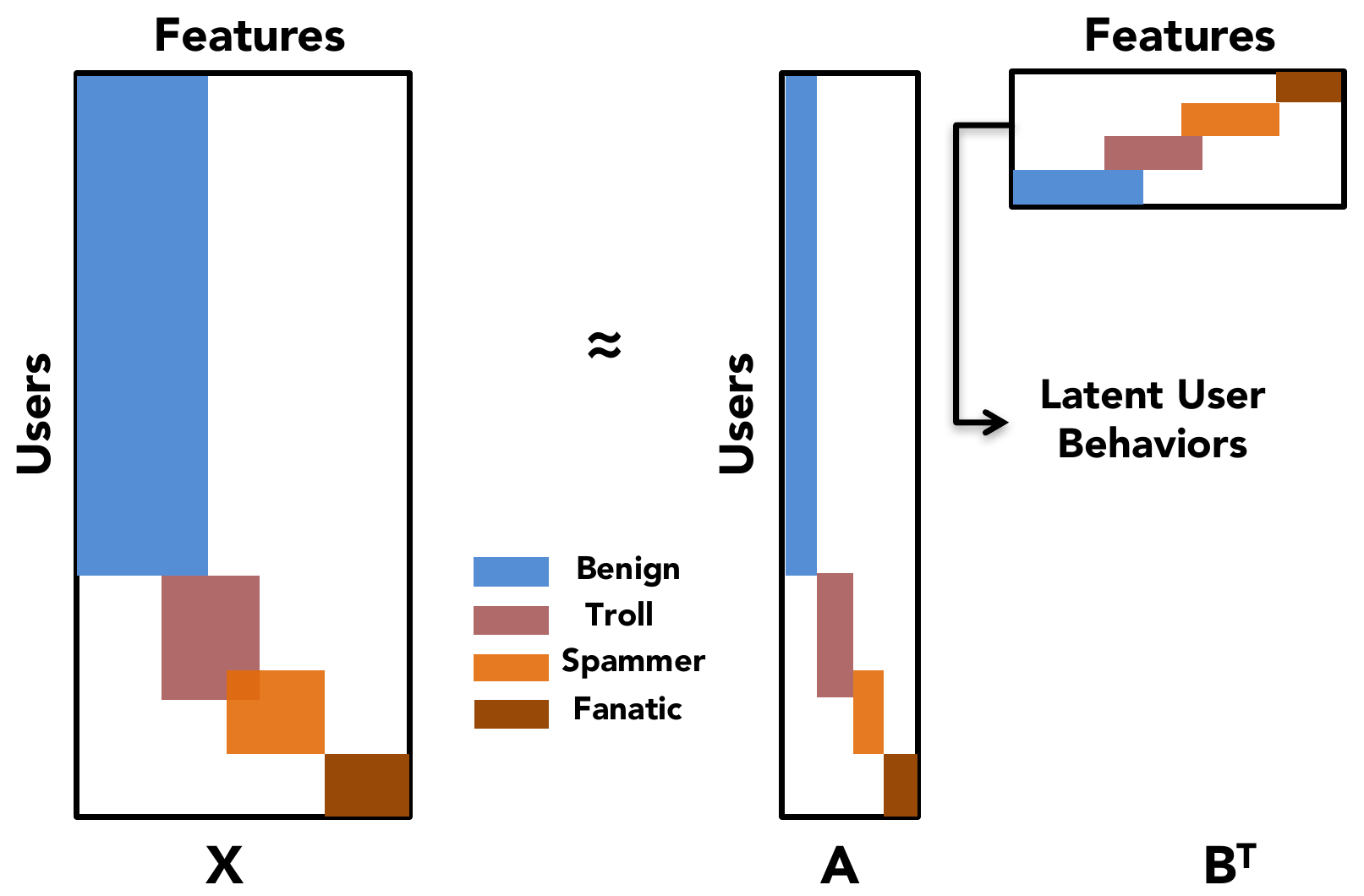}
\caption{Co-clustering discovers latent user behaviors and the assignment of users and features to those behaviors.}
\label{fig:coclustering}
\end{center}
\end{figure}

An illustrative example is shown in Figure \ref{fig:coclustering}, where we have four distinct latent user behaviors: ``Benign'', ``Troll'', ``Spammer'', and ``Fanatic''. After properly rearranging the users and the features, in this simple example we see that each latent user behavior is forming a block within the matrix. Notice that blocks can overlap since: (a) in some cases a user can be classified with different types of behavior, and (b) some features may be associated with multiple types of behavior. Mathematically, each block (formally co-cluster) in the data can be seen as (approximately) a rank-one matrix. By definition, a rank-one matrix can be written as the outer product of two vectors, i.e., $\mathbf{ab}^T$. We, thus, can use the following formulation for co-clustering, which was originally proposed in \cite{papalexakis2013k}:
\begin{equation}
	\displaystyle{\min_{\mathbf{A\geq0,B\geq0}} \| \mathbf{X} - \mathbf{AB}^T\| + \lambda \sum_{i,r}|\mathbf{A}(i,r) | + \lambda \sum_{j,r} |\mathbf{B}(j,r) | } 
\end{equation}
The above equation decomposes the data matrix $\mathbf{X}$ into a sum of rank-one matrices (each one being approximately a co-cluster) and the regularization penalties on the $\ell_1$ norms of matrices $\mathbf{A,B}$ (whose columns hold the indicator vector for each co-cluster, as shown in Fig. \ref{fig:coclustering}) ensure that the solution is sparse, so that only the users and features that belong to a particular co-cluster have a non-zero (and positive) value in our indicator vectors.
Using this type of overlapping co-clustering, a user may be assigned to multiple \lubs at the same time, with varying degrees of participation. This enhances interpretability because it can give further insight on whether e.g., a user has more than one roles or their account has been hijacked and some of their comments are benign but some are malicious.
Finally, the above co-clustering formulation is {\em lossy}, which means that it does not require all users to belong to at least one co-cluster. In other words, the above formulation focuses on identifying the most dominant latent behaviors and the users that engage in them, and leaves out users who do not clearly belong in one (or more) of those latent behaviors. We, henceforth, refer to those users as ``non-clustered''.

%% file: 502_methodology.tex
\subsection{Methodology}

\begin{figure}[ht]
	\centering
   \subfigure[]{
    	\includegraphics[width=\fsizezwei\columnwidth]{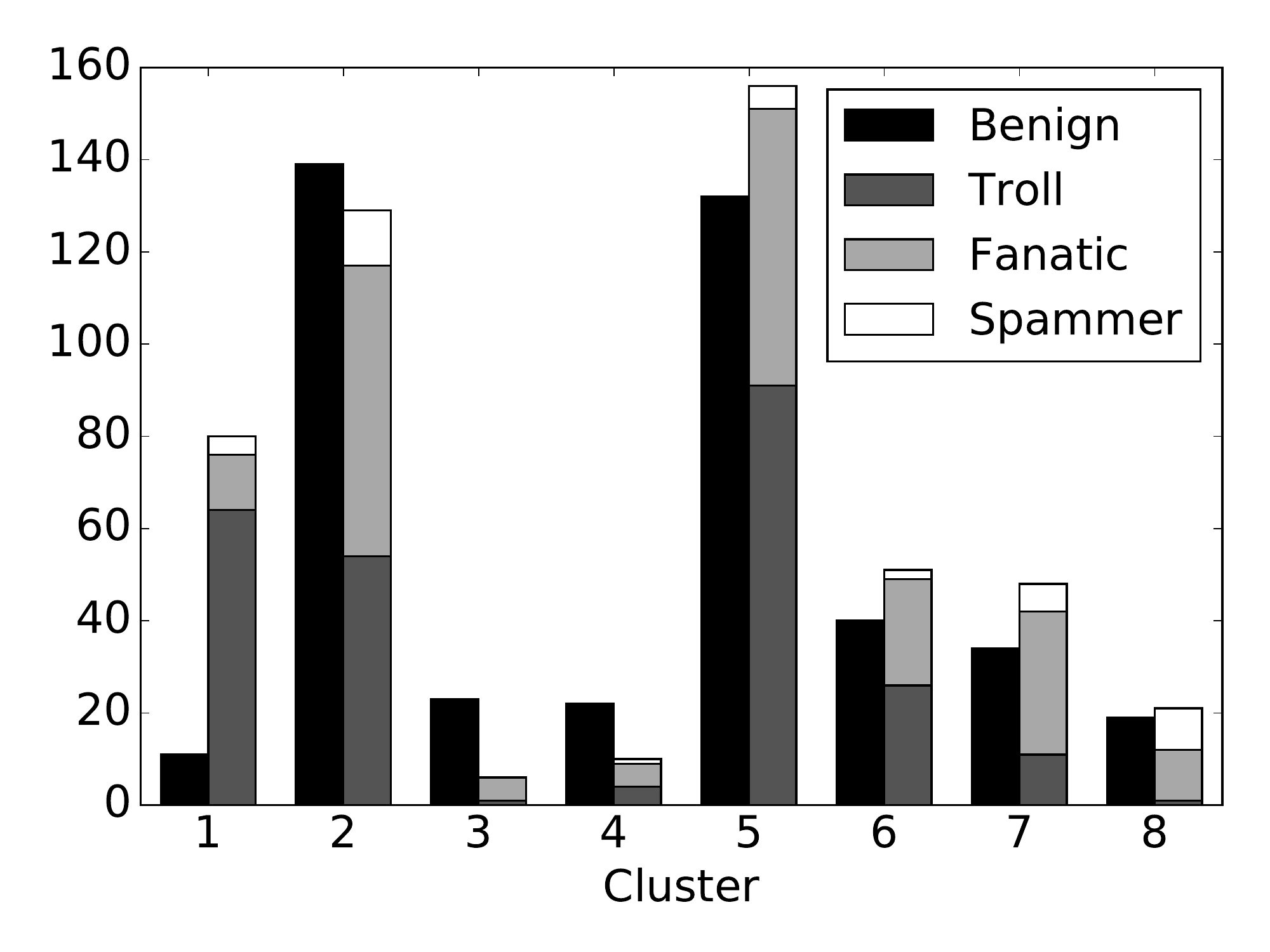}
        \label{fig:clusters}        
    }
    \subfigure[]{
    	\includegraphics[width=\fsizezwei\columnwidth]{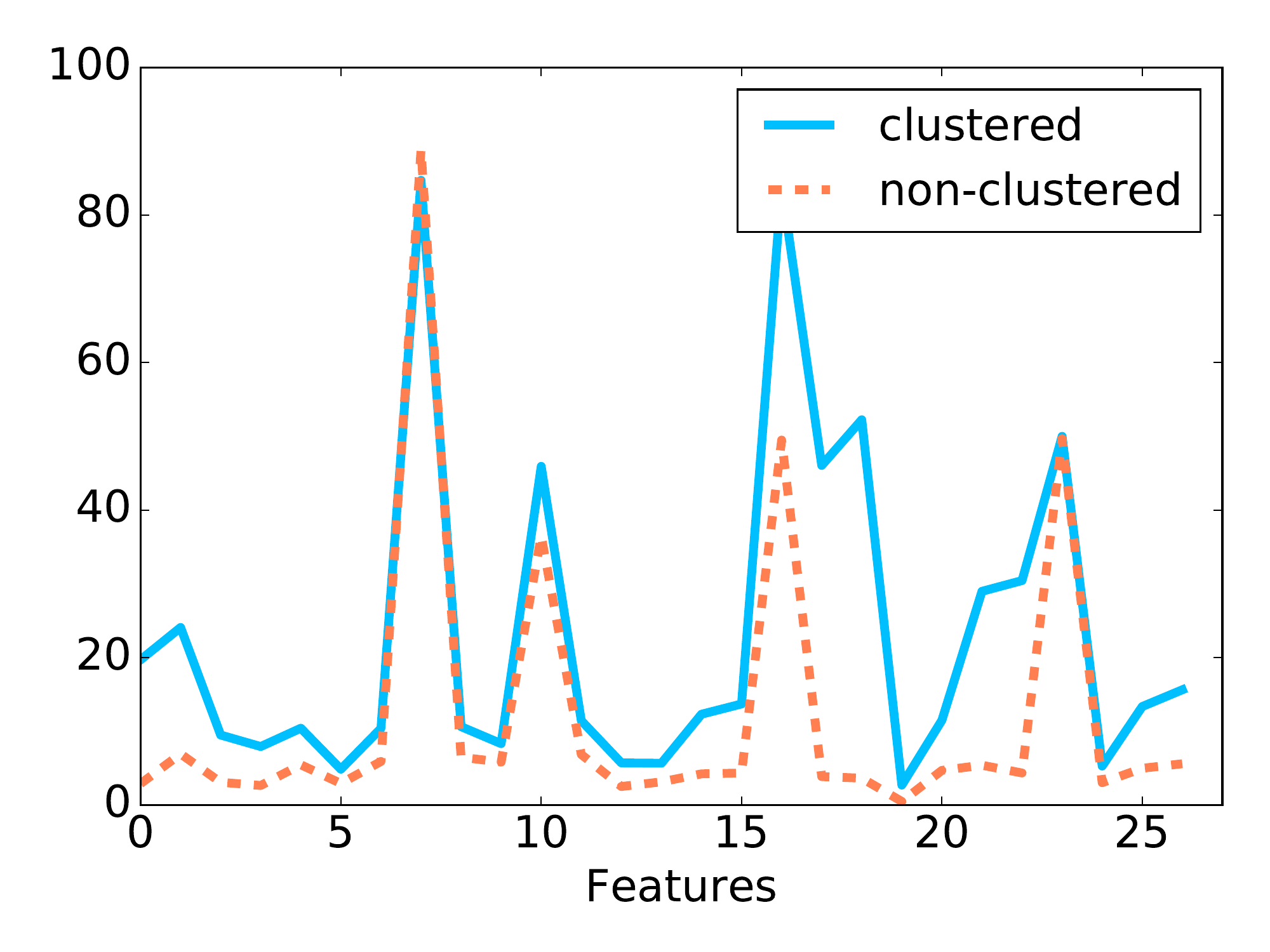}
        \label{fig:latent_behavior}
    }
   	\caption{\textbf{Co-clustering.} (a) Distribution of users in each cluster. (b) Example of \lub from cluster 1. 
}
	
\end{figure}

We apply co-clustering to the same dataset we used for \role classification. This dataset contains 200 benign users and 200 misbehaving users, which include 104 trolls, 21 spammers and 75 fanatics. After fine tuning the parameters of co-clustering, we find that the best number of co-clusters in the data is 8. An indication for this number is that the algorithm returned empty clusters for larger numbers of clusters. We show the distribution of users in each clusters in Figure~\ref{fig:clusters}. 



The size of clusters varies from 29 to 268, indicating that co-clustering could not only capture the common behaviors but also the obscure ones (small clusters). In each cluster, the clustered users will exhibit a different level of value on multiple features comparing to non-clustered users
as we present in Figure~\ref{fig:latent_behavior}. The X-axis demonstrates the 27 features used to capture the cluster and the Y-axis is the mean of features value. Clustered users have a much higher value on feature 18 and 21 which are the percentage of capital letters and number of sentences. This explains users in this cluster exhibits a \textit{\lub} of using more capital letters and more sentences in their comments.



\textbf{Using \lubs to identify misbehaving users: 0.77 AUC.} In order to show that the \lubs can be used to identify misbehaving users, we take \textbf{A} matrix from the co-clustering result. Each row in the matrix is a user and each column corresponds to a cluster, with the $(i,j)$ value being the ``membership'' of user $i$ to the cluster $j$. We use these values as  features to train a classifier. Note that for each set of features we use the classifier that yielded the highest performance. In the case of raw features this was Random Forest, but for the case of the \lub features it was Support Vector Machine (SVM) \footnote{This behavior is expected since the \lub features, contrary to the raw features, are embeddings of the users in a vector space, thus SVM should work better than Random Forest.}. We thus report results using SVM. Our result shows that we can still accurately identify misbehaving users with an AUC of 0.77 with 86.3\% precision and 63.5\% recall. The role classification result is shown in Table~\ref{tab:role_classification_coclustering}.


\textbf{Using \lubs to identify misbehaving \roles: 73.3\% overall accuracy.} Using 10-fold cross validation, we
assess the ability of our method to identify misbehaving roles, and we show the results in
Table~\ref{tab:role_classification_coclustering}. 

The overall classification accuracy when using the latent behaviors is lower than using the raw features, albeit still very good. However, we argue that the reduction in accuracy comes with an increase in the {\em interpretability} of the classification using latent behaviors, as we present below. 



\textbf{Interpreting \lubs.} 
Due to the space limitation, we are not able to discuss all \lubs in the data. We will, thus, focus on the \lubs that help us understand misbehaving users.

\begin{enumerate}
\item \Lub 1: Users who exhibit such behavior are, on average, making more comments to articles, are highly active between 7-13 UTC (midnight in US), use more capital letters and also more characters and sentences on average in their comments.
\item \Lub 3: Users are highly active between 13-26 UTC, having \textasciitilde 60\% more comments and degrees of collaboration than the average.
\item \Lub 4: Users are active between 3-8 UTC, making slightly more (24\%) comments than the average.
\item \Lub 6: Users are active between 6-14 UTC, making comments which have better readability.
\end{enumerate}

Figure~\ref{fig:clusters} shows that 87.9\% of users who exhibit \lub 1 are misbehaving users. This is in contrast to users of \lub 6, which are active at similar periods of time but only 56\% of them are misbehaving users. This explains that using more capital letters and making longer comments are the characteristic of misbehaving users.

Another interesting observation here is that, the active period could also imply whether a user is misbehaving or not. We see that there is an active time shift from \lub 3, \lub 4 to \lub 6, and the percentage of misbehaving users are 20.7\%, 31.3\% and 56\% respectively. If we assume that most of the users are active in the timezone between ET and PT, then we may further deduce that the later at night a user is active, the more likely they are to be misbehaving.




\begin{table}
	\tsize
    \renewcommand{\arraystretch}{1.3}
    \centering
    \caption{Role classification result (co-clustering)}
    \label{tab:role_classification_coclustering}      
	\begin{tabular}{p{2.5cm}|p{2cm}|p{2cm}}
    \hline
    \textbf{Role} & \textbf{Precision} & \textbf{Recall} \\
    \hline
    Trolls & 78.7\% & 71.2\% \\
    Spammers & 60.7\% & 81\% \\
    Fanatics & 76.4\% & 73.3\% \\
    Benign & 71.4\% & 73.5\%\\
    \hline
    \textbf{Overall accuracy} & \multicolumn{2}{l}{73.3\%} \\
    \hline
	\end{tabular}
    \vspace{-0.1cm}
\end{table}

%% file: 600_related.tex
\section{Related Work}\label{sec:related}

We only have space to review indicative studies in five areas. 

{\bf Detecting malicious behavior}. Many studies detect anti-social behavior \cite{cheng2015antisocial}\cite{chenganyone}, bots\cite{chu2012detecting}\cite{morstatter2016new} or cyber-bullying \cite{hosseinmardi2016prediction} but they focus on OSNs. 
A recent work \cite{cheng2015antisocial} 
analyzes the comments on three websites using Disqus, but the focus is to predict if a user will be banned
due to misbehavior, which is somewhat different from ours:
(a)  some misbehaviors are not necessarily reported, as we saw,
and (b) we have a fine-grained definition of misbehavior.


{\bf Detecting malicious behavior}. Many studies detect anti-social behavior \cite{cheng2015antisocial}\cite{chenganyone}, bots\cite{chu2012detecting}\cite{morstatter2016new} or cyber-bullying \cite{hosseinmardi2016prediction} but they focus on OSNs. 
A recent work \cite{cheng2015antisocial} 
analyzes the comments on three websites using Disqus, but the focus is to predict if a user will be banned
due to misbehavior, which is somewhat different from ours:
(a)  some misbehaviors are not necessarily reported, as we saw,
and (b) we have a fine-grained definition of misbehavior.

{\bf Analyzing text of online users.}
Many studies leverage textual
analysis to detect spammers on OSNs \cite{sculley2007relaxed} and blogs 
\cite{mishne2005blocking} or how to achieve the same objective 
by analyzing the  behavioral patterns \cite{sureka2011mining}.

{\bf Modeling and understanding online user behavior.}  Several works study temporal and behavioral patterns of OSN users  \cite{ferraz2015rsc} \cite{devineni2015if} \cite{gu2016web}. 
Some recent studies also use mobile phones and OSNs to profile users' psychological
states \cite{wang2014studentlife}\cite{de2014characterizing}.

%% file: 700_conclusion.tex
\section{Conclusion}\label{sec:conclusion}

We develop a systematic and comprehensive methodology to identify malicious users on commenting platforms.
The novelty of our work lies in its: (a) fine-grained classification of malicious behavior,  and (b) interpretable approach to classification.
From an algorithmic point of view our work uses a comprehensive set of 73 features across four different types of information, and a novel two-stage classification.

The overall classification accuracy of our approach is 
80.8\% for the fine-grained 4-class problem.
We also identify several unusual and surprising behaviors, and
we provide a broader understanding of the users of these platforms.

Our work is a first step towards an efficient, easy-to-understand and easy-to-manage approach to safeguarding commenting platforms from parasitic and malicious users.

In the future, we aim at: (a) create a public larger labeled datasets to facilitate further research, 
(b) consider more categories of malicious \roles,
and (c) conduct large scale study of the evolution of malicious behavior on commenting platforms.
\vspace{-0.3cm}